\documentclass[aip,jcp,twocolumn,groupedaddress]{revtex4}
\usepackage{epsfig}
\usepackage{color}
\usepackage{amsmath}

\begin{document}
\def\br{{\bf r}}
\def\vcmi{v_{cm,i}}

\title{Analytical Rescaling of Polymer Dynamics from Mesoscale Simulations}
\author{I. Y. Lyubimov} 
\author{J. McCarty}
\author{A. Clark}
\author{M. G. Guenza\footnote{Author to whom correspondence should be addressed. Electronic mail: mguenza@uoregon.edu}}
\affiliation{Department of Chemistry and Institute of Theoretical Science,
University of Oregon, Eugene, OR 97403, USA}

\begin{abstract}
We present a theoretical approach to scale the artificially fast dynamics of simulated  coarse-grained polymer liquids down to its realistic value. As coarse-graining affects entropy and dissipation, two factors enter the rescaling: inclusion of intramolecular vibrational degrees of freedom, and rescaling of the friction coefficient. Because our approach is analytical, it is general and transferable. Translational and rotational diffusion of unentangled and entangled polyethylene melts, predicted from mesoscale simulations of coarse-grained polymer melts using our rescaling procedure, are in quantitative agreement with united atom simulations and with experiments.
\end{abstract}

\maketitle

\section{Introduction}
The development of a systematic approach to bridge time scales between different hierarchical levels of description is an important goal in many areas of the physics of complex systems.\cite{GellMann} Furthermore, understanding the dynamics of polymeric liquids is relevant for many technological applications. Polymeric materials are processed in their liquid state, and the custom tailoring of new materials requires detailed predictions of their mechanical and dynamical properties to be made based on their chemical structure. To this end, molecular dynamics (MD) simulations shed light on the properties of complex systems; however, MD is limited by the precision of the calculations, which degrades with the number of computer iterations.\cite{shadowing} Quantitative predictions of the dynamics of unentangled polymer liquids are possible through simulation runs that are computationally demanding because the polymer diffusion coefficient, $D$, scales with the degree of polymerization, $N$, as $D\propto N^{-1}$. Even more demanding are simulations of liquids of long, entangled, polymeric chains, where the diffusion coefficient scales as $D\propto N^{-2}$. For entangled polymer liquids it is therefore difficult to obtain well-equilibrated samples or to simulate the system for several relaxation cycles, which would improve the precision of the calculated time-correlation functions. 

To reach the long time and length scales of interest, special strategies need to be employed. For example, some gain in computational time has been achieved by speeding up the equilibration process  through an end-bridging Monte Carlo algorithm.\cite{Mavran} Moreover, if only qualitative, and not quantitative, predictions are sought, it is possible to adopt simplified intra- and intermolecular potentials, which can speed up each simulation step,\cite{Hess,Likhtman} or purely phenomenological coarse-grained descriptions, which reproduce the expected dynamical scaling behavior.\cite{Likhtman,entangleds} However, quantitative computational methods that precisely relate chemical structure to the long-time properties of the polymeric liquid are \mbox{still lacking.}

There is a need for computationally efficient, predictive methods to simulate polymer liquids and complex fluids in the long-time regime. The strategy we are pursuing here is to develop coarse-graining methods and dynamical rescaling procedures which start from first principles theory.\cite{Ottinger,YAPRL} The goal is to obtain quantitative predictions of real polymer dynamics directly from properly rescaled fast mesoscale (MS) simulations of coarse-grained liquids. 

In a coarse-grained description, the system is specified by a set of relevant mesoscopic variables while smaller length scale variables are omitted.\cite{MSMD} This is done at the expense of entropy and dissipation (friction), which are underestimated by coarse-graining.\cite{Ottinger} Because of the reduced molecular degrees of freedom and the simplified energy landscape, MS-MD simulations require less computational time than atomistic simulations of the same systems.\cite{YAPRL,DePablo,Klein,M-Plathe} However, because the effective energy landscape of the coarse-grained representation is artificially smooth, MS-MD simulations predict accelerated dynamics, which need to be rescaled to produce realistic values. 

The enhanced diffusion in coarse-grained systems arises from the ``soft" nature of the intermolecular potential. This is advantageous in enabling larger time steps to be used in integrating the equations of motion, and an efficient sampling of the energy landscape, which leads to good statistical averages of the structural properties on the large scale.\cite{Nielsen} However, the measured dynamics of coarse-grained systems is too fast, and as of yet, there is no clear procedure to quantitatively re-scale these accelerated dynamics.

In an effort to develop quantitative rescaling methods, it is custom to build a  numerical ``calibration curve" obtained from direct comparison of MS-MD time correlation functions with atomistic simulations. The ``calibration curves" are parametric, normal mode dependent, and specific to the system against which they are optimized, as well as to its thermodynamic conditions. Building these parametric curves in part defeats the purpose of the coarse-graining procedure as it requires running many atomistic simulations to optimize the fitting parameters. If the level of coarse-graining is low, i.e. if the average is performed over a small number of atoms, the correction to the dynamics is only perturbative and the parametric rescaling works well.\cite{Kremer} This explains the success of united-atom (UA) simulations.\cite{MONDE,Mavran} However, because the gain in computational time increases with the level of coarse-graining, simulations of slightly coarse-grained systems,e.g. UA-MD, afford a limited gain in time and length scales.

The most gain in computational efficiency is achieved through large-scale coarse-graining, as the one adopted in this paper. This is most useful when studying bulk physical quantities, e.g. viscosity, or systems with large-scale fluctuations, e.g. approaching spinodal decomposition. \cite{macromol} Once simulations of heavily coarse-grained systems are combined with local scale simulations in a multiscale procedure, they provide the complete description of the system at all lengthscales of interest.\cite{multis}

This paper presents a derivation of a first-principles approach to rescale the dynamics from MS-MD simulations of a coarse-grained polymer melt to the values of an atomistic description. The rescaling is analytical, general and transferible. The favorable comparison with atomistic simulations and experimental data in different thermodynamic conditions supports the validity of the proposed procedure. The theoretical basis of the rescaling rests on the fact that the accelerated dynamics is a result of the missing dissipation due to the eliminated degrees of freedom.

The paper is organized as following. In Section 2 we briefly review our coarse-grained model, while the coarse-grained potential and mesoscale simulations are described in Section 3. In the following section we present our atomistic model and the calculation of the energy due to the internal degrees of freedom. Rescaling of the friction coefficient is discussed in Section 5, followed by a comparison of the diffusion coefficients predicted from MS-MD simulations after rescaling, with united-atom simulation and experimental data. A brief discussion concludes the paper.

\section{Coarse-grain model}
Our coarse-grained representation models polymers as interacting soft-colloidal particles with repulsive interaction of the order of the size of the macromolecule, defined by the radius-of-gyration, $R_g$.\cite{YAPRL}  The total distribution function is derived from the solution of a generalized Ornstein-Zernike equation, treating coarse-grained sites as auxiliary sites, and atomic sites as real sites\cite{K2002}. 
In reciprocal space, the total distribution function is 

\begin{eqnarray}
\label{eqnuno}
h(k) = \left[\omega^{cm}(k)/\omega^{mm}(k)\right]^2 h^{mm}(k) \ ,
\end{eqnarray}

\noindent where the superscript ``mm" identifies the  monomer-monomer distribution, while ``cm" indicates the distribution of monomers with respect to the center-of-mass. Eq.(\ref{eqnuno}) is solved by assuming a Gaussian description of the intramolecular site distribution, which is an accepted approximation for polymers in a liquid state. The form factors entering Eq.(\ref{eqnuno}) are approximated by $\omega^{cm}(k)=N e^{-\frac{k^2 R_g^2}{6}}$ and $\omega^{mm}(k) = N/(1+k^2 R_g^2/2)$, which is the Pad\'e approximant of the Debye function $\omega^{mm}(k) = 2 N (e^{-k^2 R_g^2} + k^2 R_g^2 - 1)/(k^4 R_g^4)$. For the monomer-monomer total intermolecular correlation function $h^{mm}(k)$, we use the thread-limit polymer reference interaction site model description,\cite{prism} in which $h^{mm}(k) = h_0/(1+\xi_\rho^2 k^2)(1+k^2 \xi_c^2)$. Here, $\xi_\rho$ is the length scale of density fluctuations defined as $\xi^{-1}_\rho=\xi^{-1}_c +\xi'^{-1}_\rho$, with $\xi_c=R_g/\sqrt{2}$ the length scale of the correlation hole, and $\xi'_\rho=R_g/(2\pi\rho^*_{s})$ with $\rho^*_{s}=\rho_{s}R^3_g$ being the reduced molecular number density. The number density of soft colloidal particles $\rho_{s}=\rho/N$ with $\rho$ being the monomer density and $h_0=(\xi_\rho^2/\xi_c^2 -1)/\rho_{s}$. 
 
The structure of the coarse-grained polymer liquid is described by the total distribution function,\cite{YAPRL} approximated for  polymer chains with $N\ge 30$ as
\begin{equation}
\begin{split}
\label{EQ:HCCA}
h(r,\xi_\rho)  \approx 
&-\frac{39}{16}\sqrt{\frac{3}{\pi}}\frac{\xi_\rho}{R_g}
\left(1+\sqrt{2}\frac{\xi_\rho}{R_g}\right) \\
 &\times \left[1-\frac{9r^2}{26R_g^2} + \mathcal{O}\left(\frac{\xi_{\rho}^2}{R_g^2},\frac{{r}^4}{R_g^4}\right)\right] e^{-\frac{3 r^2}{4R_g^2}} \, ,
 \end{split}
 \end{equation}

\noindent which results from the analytical Fourier transform of $h(k)$. The structure of the liquid on length scales of the order of the polymer radius-of-gyration and larger, is well described by Eq.(\ref{EQ:HCCA}), which is in quantitative agreement with both atomistic and coarse-grained simulations.\cite{YAPRL} The coarse-grained description of Eq.(\ref{EQ:HCCA}) is thermodynamically consistent with the atomistic representation, e.g. of the liquid compressibility.\cite{YAPRL} Because the total correlation function of the coarse-grained representation is analytical, i.e. it depends explicitly on density and molecular parameters, it is also general and state-point transferable.

\section{Coarse-grained Potential and Mesoscale Simulations}
Each soft-colloidal particle interacts with other colloids through an effective potential of the range of the overall polymer dimension, $R_g$. While hard-sphere systems are best described by a Percus-Yevick closure, the Hyper-Netted Chain (HNC) closure works best for systems with soft potentials,\cite{mcquarrie}  including the mesoscopically coarse-grained polymer melts investigated here,\cite{YAPRL,edjcp} and polymer coils in dilute or semidilute solutions.\cite{Louis} The HNC potential between a pair of coarse-grained units is derived by applying the closure $\beta v(r) = h(r) - ln[h(r)+1]-c(r)$, where the direct correlation function is defined by the Ornstein-Zernike relation in reciprocal space $c(k) = h(k)/(1+\rho_s h(k))$.\cite{SMPLQ} 

The potential between two spheres is calculated numerically, after Fourier transform, from the analytical expression, Eq.(\ref{eqnuno}), by adopting the Debye form of the monomer intramolecular distribution, $\omega^{mm}(k)$. The Debye approximation has been shown to better represent simulation data  than its Pad\'e approximant.\cite{edjcp}  Tabulated HNC potentials are input to the mesoscale simulations. 

MS-MD simulations  of  polymer liquids are performed in the microcanonical ensemble, where each molecule is represented as an interacting soft-colloidal particle. In the initialization step, all particles are placed on a lattice with periodic boundary conditions. Each site is given an initial velocity and subsequently, the system is evolved using a velocity Verlet integrator.  Equilibrium is induced in the ensemble by rescaling the velocity at regular intervals until the desired average temperature is reached. At this stage, velocity rescaling is discontinued and trajectories are collected over a traversal of $\sim8R_g$, while the temperature is monitored to assure that it fluctuates around the desired equilibrium value. Because of the form of Eq.(\ref{EQ:HCCA}), the  simulation uses reduced quantities of distance, $R_g=1$, mass, $m=1$, and energy, $k_BT=1$.

A typical MS-MD simulation for our model is performed  on a single-CPU workstation, and consists of $\sim3000$ polymers, evolving for a duration of $\sim4$ hours. The MS-MD simulation provides identical structural information to that of the analogous atomistic or united-atom (UA) simulations on length scales equal and larger than the polymer $R_g$. However, the MS-MD requires a much smaller computational power than the atomistic  and UA-MD simulation, which is typically performed on a liquid of $\sim 500$ polymers, on a parallel supercomputer. The convenient requirements in computational power of the MS-MD simulation allows one to increase considerably the number of particles and the simulation box size without dramatically affecting the computational time, thus improving the precision on the large-scale data collected.

\section{Mapping of the atomistic description onto a freely-rotating-chain model}
The correction in Helmhotz free energy, which accounts for the discarded internal degrees of freedom, is calculated starting from the atomistic representation of the liquid, where each chain is described as a collection of beads connected by springs defined by an effective intramolecular quadratic potential $U(r)=3k_BT/(2l^2)\sum_{i,j=1}^N \textbf{A}_{i,j} \textbf{r}_i \cdot \textbf{r}_j $. Here $\bf{A}$ is the connectivity matrix, which represents the structure and local flexibility of the polymer, $\textbf{r}_i$ the position of unit $i$ in a chain of $N$ beads, and $\textbf{l}_i=\textbf{r}_{i+1}-\textbf{r}_i$ the bond vector connecting two adjacent beads.\cite{jpcmreview} For polyethylene melts, each polymer is represented as a freely-rotating-chain (FRC), finite in size, with semiflexibility parameter $g=\left< (\textbf{l}_i \cdot \textbf{l}_{i+1})/(l_i l_{i+1}) \right> =0.785$ and fixed bond length $l=1.54$\AA. This bead-and-spring model of the FRC has been shown to represent correctly the dynamics of polyethylene melts as measured in united-atom simulations\cite{manychain} and in experiments.\cite{Richter} A FRC model was also successfully adopted to model the dynamics of polymer melts with different molecular architectures,\cite{jpcm} and even proteins,\cite{jpcmreview,protein} once the proper semiflexibility parameters are selected.    
 
To map the MS-MD simulation onto a real system, the reduced unit of time needs to be properly rescaled. Because energy is dissipated in internal degrees of freedom in the atomistic representation, the contribution due to the internal vibrational modes is included in the coarse-grained representation by rescaling the time in the MS-MD simulation, $\tilde{t}$, by  the amount of internal free energy dispersed in vibrational modes in the atomistic description as $t=\tilde{t} R_g \sqrt{3 mN /(2 k_BT)}$, with the particle mass, $m$, and size $R_g$. By rescaling the internal free energy, we accounts for the change in entropy in the coarse-grained description. In the next section we derive the friction rescaling.

\section{Rescaling of the friction coefficient}
To account for the change in dissipation caused by coarse-graining, we start from the diffusion coefficient measured in the MS-MD simulation, $D^{MS}_{t}$, and we derive the center-of-mass (cm) diffusion coefficient of the polymer, $ D_{cm}$, through the rescaling of the friction as 

\begin{equation}
D_{cm}=D^{MS}_{t} \zeta_{\text{s}}/(N \zeta_m) \, .
\end{equation}

The correction factor is calculated from the ratio of the cm friction in the soft colloid representation $\zeta_{s}$ and the cm friction in the atomistic representation $N \zeta_m$, with $\zeta_m$ the monomer  friction. Each friction coefficient is evaluated by solving the memory function in the corresponding Generalized Langevin Equation. These memory function definitions result from the straightforward application of Mori-Zwanzig projection operators to the Liouville equation, where either the monomer (atomistic description) or the center-of-mass (soft colloid description) of the polymer are assumed to be the ``relevant slow variables."\cite{jpcmreview,SMPLQ}

In the soft colloid representation the friction coefficient is defined as 

\begin{equation}
\begin{split}
\zeta_{s} \cong &\frac{\beta}{3} \rho_{s} \int_0^{\infty} d t 
  \int d\mathbf{r} \int d\mathbf{r'} g(r) g(r') F(r) F(r') \mathbf{\hat r} \cdot \mathbf{\hat r'}\\ 
  &\times \int d \mathbf{R} S(R;t) S(|\mathbf{r}-\mathbf{r'}+\mathbf{R}|;t) \, , 
\label{EQ:MFCM}
\end{split}
\end{equation}

\noindent where $\beta = 1/k_B T$, $g(r)=h(r)+1$ is the radial distribution function, $F(r)= \beta^{-1} (d \ln g(r)/dr)$ is the total force exerted by the surrounding fluid on the colloid, and $S(k)=1+\rho_{s} h(k)$ is the structure factor of the fluid surrounding the colloid. The unit vectors  $\mathbf{\hat r}$ and $\mathbf{\hat r'}$ define the direction of exerted forces.

In Eq.(\ref{EQ:MFCM}), the projected dynamics  has been substituted with the real (unprojected) dynamics, which is a valid approximation when the Langevin equation is expressed as a function of slow variables for the diffusive regime.\cite{SMPLQ} In the long-time regime, the relaxation of the liquid is dominated by the polymer center-of-mass diffusion, $D$, here represented by the cm of the colloidal particle. In Fourier space the dynamic structure factor of the liquid reads $S(k;t)\approx S(k) exp({- k^2 D t})$. Evaluating the integral with use of Eq.(\ref{EQ:HCCA}) gives

\begin{equation}
\begin{split}
D \beta \zeta_{s} \approx &4 \sqrt{\pi} \rho_{s} R_g \xi_\rho^2 \left(1+\frac{\xi_\rho}{\xi_c} \right)^2 \frac{507}{512}\\ 
&\times \left[ \sqrt{\frac{3}{2}}  + \frac{1183}{507} \rho_{s} h_0 + \frac{679\sqrt{3}}{1024} \rho_{s}^2 h_0^2 \right] \, .
\label{EQ:zeta_soft_mean_force}
\end{split}
\end{equation}

Eq.(\ref{EQ:zeta_soft_mean_force}) expresses the friction coefficient of a soft colloidal particle.

In the atomistic description, the monomer friction coefficient is defined  by the memory function as

\begin{equation}
\begin{split}
\zeta_{m}  \cong &\frac{1}{N} \sum_{j, i=1}^{N} \int_0^{\infty} d \tau \frac{\beta}{3} \rho \int d\mathbf{r} 
\int d\mathbf{r'} g(r) g(r') F(r) F(r') \\   
&\times\mathbf{\hat r} \cdot \mathbf{\hat r'} \int d \mathbf{R} S_{i,j}(R;t) S(|\mathbf{r}-\mathbf{r'}+\mathbf{R}|;t) \, , 
\label{EQ:MFMON}
\end{split}
\end{equation}

\noindent where the dynamic structure factor of the surrounding liquid is approximated as $S(k;t)\approx S(k) exp({- k^2 D t})= \left[\omega(k) + \rho h(k)\right] exp({- k^2 D t})$, with $\omega(k)$ being the intramolecular static structure factor. This expression for $S(k;t)$ assumes that in the long time regime  the relaxation of the liquid is dominated by the polymer center-of-mass diffusion, consistently with the soft-colloid representation.

To evaluate Eq.(\ref{EQ:MFMON}), we approximate the potential as an effective hard-core potential with a diameter $d$ to be defined.\cite{SMPLQ} In hard-core fluids, $g(r)F(r)=g(d)\beta^{-1}\delta(r-d)$. Working in reciprocal space, the integrals in Eq.~(\ref{EQ:MFMON}) can be performed analytically to give an expression for the dynamical quantity $D \beta \zeta_m$ depending on two length scales: $R_g$, and $d$. The result is lengthy, and it is not reported here.\cite{note} The values of $R_g$ used are the ones reported in Tables \ref{TB:SimParam} and \ref{TB:ExpParam}. 

\begin{table}[h!b!p!]
\caption{MS-MD Parameters - UA-MD data}
\centering
\renewcommand{\baselinestretch}{1}\normalsize
\begin{tabular*}{230pt}{@{\extracolsep{\fill}}lrccc}
  \hline \hline
Polymer & N & T [K] & $\rho$ [sites/\AA$^3$] & $R_{g}$ [\AA] \\
\hline 
PE30$^a$    &   30  &   400   &   0.0317    &    7.97    \\
PE44$^a$    &   44  &   400   &   0.0324    &   10.50    \\
PE48$^b$    &   48  &   450   &   0.0314    &   10.54    \\
PE66$^a$    &   66  &   448   &   0.0329    &   13.32    \\
PE78$^b$    &   78  &   450   &   0.0321    &   14.35    \\
PE96$^a$    &   96  &   448   &   0.0328    &   16.79    \\
PE142$^b$   &   142 &   450   &   0.0327    &   20.51    \\
PE174$^b$   &   174 &   450   &   0.0328    &   22.92    \\
PE224$^b$   &   224 &   450   &   0.0329    &   26.28    \\
PE270$^b$   &   270 &   450   &   0.0330    &   29.27    \\
PE320$^b$   &   320 &   450   &   0.0330    &   31.31    \\
\hline \hline
\multicolumn{5}{l}{$^a$ data from ref. \cite{MONDE}; $^b$ data from ref. \cite{Mavran}}
\end{tabular*}
\label{TB:SimParam}
\end{table}

\begin{table}[h]
\centering
\caption{MS-MD Parameters - Experimental Data [$T=509$K, $\rho=0.0315$ [sites/\AA$^3$]. Data from Ref.\cite{Richter}]}
\bigskip
\begin{tabular*}{200pt}{@{\extracolsep{\fill}}lrc}\hline \hline
\mbox{Polymer}      
& $N$ & $R_g^{FRC}$[\AA] \\
\hline
PE36   &    36   &  10.07     \\
PE72   &    72   &  14.82     \\
PE106  &   106   &  18.20     \\
PE130  &   130   &  20.25     \\
PE143  &   143   &  21.27     \\
PE192  &   192   &  24.77     \\
PE242  &   242   &  27.88     \\
\hline \hline
\end{tabular*}
\label{TB:ExpParam}
\end{table}

The monomer hard-core diameter, $d=2.1$\AA, is identical for all the samples, and is obtained by reproducing the scaling with $N$ of an unentangled melt, $D \beta \zeta_m=1/N$, for the PE 44 sample. This sample is chosen because its degree of polymerization is smaller than the entanglement one, $N_e=130$, and it is large enough to ensure Gaussian chain statistics. Once $d$ is defined, it is not changed for any other system considered, either unentangled or entangled. This is the only parameter that has to be fixed in our approach. Theoretically predicted values for unentangled systems recover the correct scaling behavior as $N D \beta \zeta_m\approx 1$. For entangled systems, we solve Eq.(\ref{EQ:MFMON}) by including a one-loop perturbation of the diffusion coefficient. For these entangled systems $N D \beta \zeta_m \propto N^{-1}$, in agreement with the known scaling behavior.

\section{Comparison of predicted diffusive dynamics with simulations and experiments}
To test our approach we compare the rescaled dynamics, predicted from mesoscale simulations, with experiments and UA-MD simulations. Each sample that we investigate is in well-defined thermodynamic conditions of density and temperature, and has a specific radius of gyration. Those quantities enter as an input to our mesoscale simulation, and also in the expressions for the rescaling of the energy and friction coefficient (see Tables \ref{TB:SimParam} and \ref{TB:ExpParam}). 

Comparison with simulation data are limited here to United-Atom simulations, but our theory is general and comparison could be made with atomistic simulations as well. The UA-MD simulations reported in this paper cover a regime from unentangled,\cite{MONDE} to slightly entangled dynamics (two entanglements per chain).\cite{Mavran} We use as an input of our approach the radius of gyration, as measured in each simulation. These values are very close to the theoretical $R_g$ values calculated using a FRC model with semiflexibility parameter $g=0.785$.

The experimental samples considered in this paper\cite{Richter} cover a region at the crossover from unentangled to entangled dynamics comparable to the one in UA-MD simulations. However, the values of the radius-of-gyration for those samples are not known, because only the degree of polymerization is reported. For those samples we assume as input values of $R_g$ those calculated using a FRC approach.

\begin{figure}[h]
 \centering 
 \includegraphics[scale=0.8]{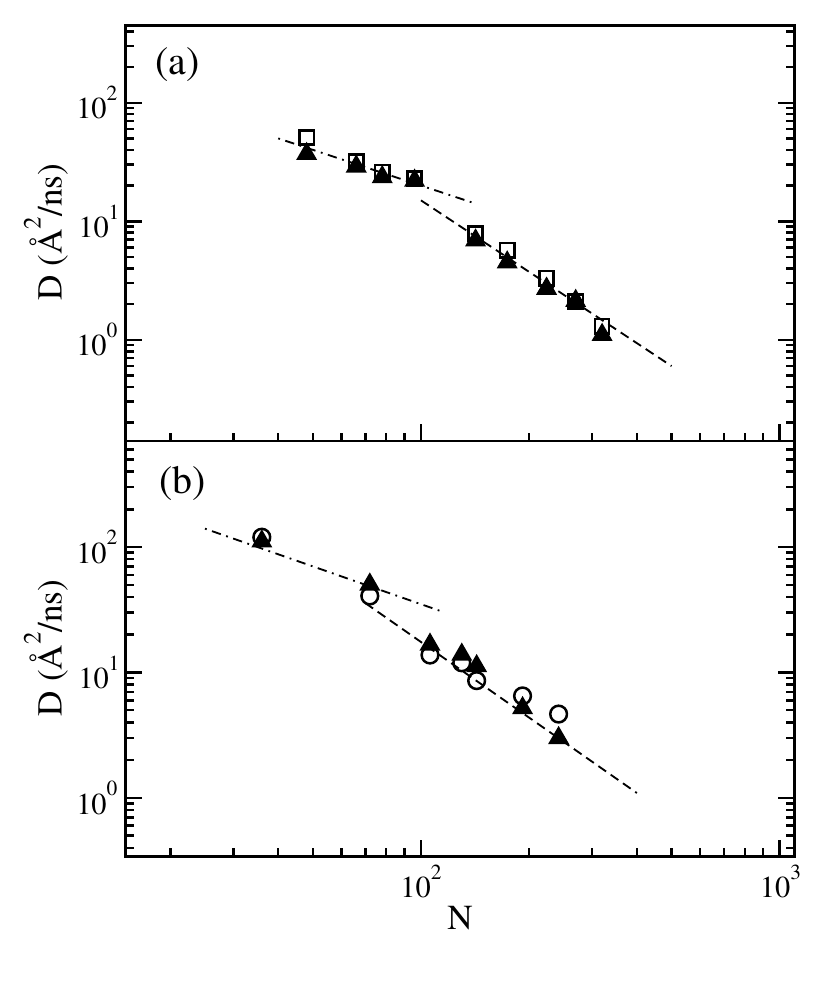}
 \caption[diffusion] {Plot of diffusion coefficients as a function of degree of polymerization, $N$. Comparison between the theoretically predicted values (triangle), simulations (square) from \cite{Mavran,MONDE}, and experiments (circle) from \cite{Richter} and references therein. Also shown is the scaling for unentangled, $N^{-1}$ (dot-dashed line), and entangled systems, $N^{-2}$ (dashed line)}
\label{FG:1}
\end{figure}

The predicted cm diffusion coefficient of a polymer chain, $D_{cm}$ is compared in Figure \ref{FG:1} against the data from simulations,\cite{Mavran,MONDE} and from experiments.\cite{Richter} We also show, as a guide to the eye, lines with the scaling behavior of unentangled and entangled systems. The agreement between calculated and measured diffusion coefficients is good over a range of the degree-of-polymerization, covering unentangled as well as entangled polymer melts. Because each simulation and experimental value is taken in slightly different thermodynamic conditions, the data points do not perfectly align along the lines of the scaling exponents in the figure. However we observe a good agreement between predicted theoretical values and measured ones in simulations or experiments.

\begin{figure}[h]
\centering
\includegraphics[scale=0.7]{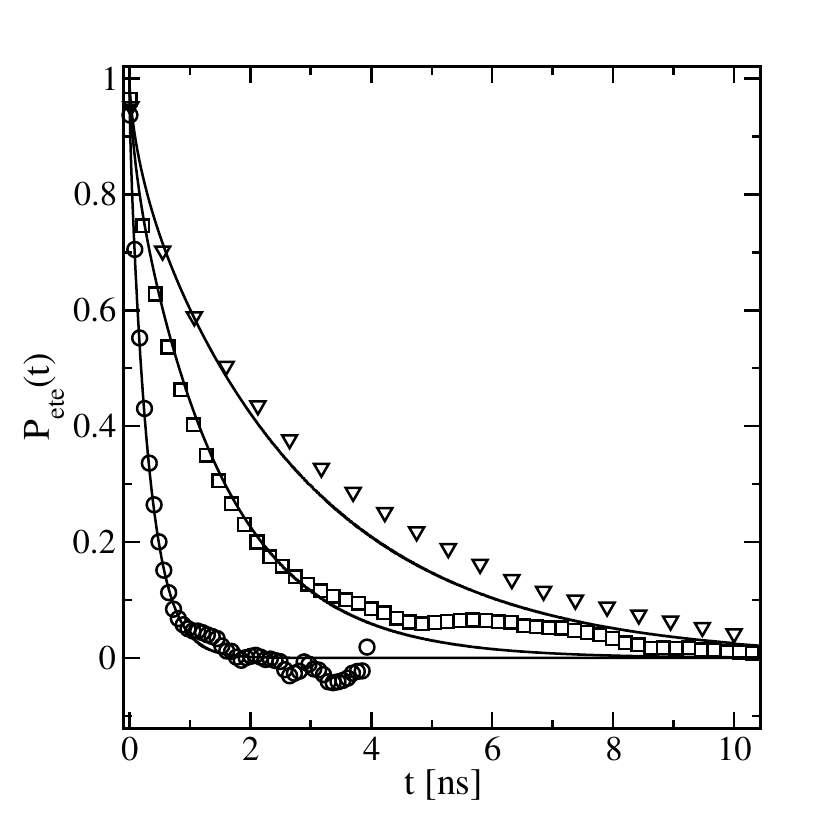}
\caption[]{Normalized rotational time decorrelation function for the end-to-end vector for the semiflexible chains with rescaled friction (solid lines) compared against simulations (symbols) for polyethylene melts of increasing length; $N=30$ (circles), $N=66$ (squares), $N=96$ (triangles)}
\label{FG:2}
\end{figure}

To test further the validity of our procedure, we calculate the decay of the rotational time-correlation function for the molecular end-to-end vector with input parameters for polyethylene and the rescaled monomer friction coefficient $\zeta=k_B T/(N D_{cm})$, and compare it against UA-MD simulations (see Figure \ref{FG:2}). Predicted and measured decays are in excellent agreement for the unentangled samples, suggesting that the proposed procedure holds for different normal modes of motion.

\section{Conclusions}
Mesocale simulations of coarse-grained systems are becoming increasingly important, as they are computationally efficient and allow for the study of systems on larger length and time scales than their atomistic counterparts. However, while structural properties on large length scales are well described by MS-MD simulations, the dynamics is unrealistically fast due to the simplified free energy landscape. In this paper we have presented an analytical, first-principles, approach to scale the dynamics measured in mesoscale simulations down to realistic atomistic values. The rescaling procedure takes into account the averaged internal degrees of freedom and enhanced dissipation due to the coarse-grainining procedure. The agreement of predicted long-time dynamics with data from simulations and experiments is quantitative. Whereas previous efforts at dynamical rescaling have used numerical calibration curves, which are specific of the system under study, our approach is analytical and thus general and transferable: it is readily applicable to systems with different thermodynamic parameters and to polymer chains of increasing degree of polymerization crossing from the unentangled to the entangled regime. The development of a general scheme to rescale the dynamics from MS-MD simulations promises to be useful in multiscale modeling techniques and fast equilibration methods employed in computer simulations of complex fluids. 

In summary, the development of schemes to rescale the dynamics from MS-MD simulations will certainly be beneficial in the advancement of multiscale modeling techniques of complex fluids. Equilibrium and non equilibrium simulations of polymer melts with different architectures should be natural implementations of the approach for future work. It should also be possible to extend this model to rescale systems represented at an intermediate level of coarse-graining as collections of soft colloidal beads, so that the internal dynamics of entangled chains can be simulated.

\section*{ACKNOWLEDGNEMTS}
We acknowledge support from the National Science Foundation. Simulation trajectories for the UA-MD simulations were kindly provided by Gary G. Grest and Vlasis G. Mavrantzas. We thank Glenn T. Evans for the careful reading of the manuscript and helpful suggestions.


\begin{thebibliography}{99}
\bibitem{note} The mathematical notebook with the solution is available upon request.
\end{thebibliography}
\end{document}